\documentclass[twocolumn,fleqn,natbib]{svjour2}
\smartqed  

\usepackage{graphicx}

 \usepackage{mathptmx}      
\usepackage{amsmath}

\journalname{Biological Cybernetics}

\begin{document}
\title{Output Stream of Binding Neuron with Feedback}
\titlerunning{Output Stream of Binding Neuron}
\author{Alexander K. Vidybida}

\institute{A.K. Vidybida \at Bogolyubov Institute for Theoretical Physics\\
Metrologichna str. 14-B, 03680 Kyiv, Ukraine\\
\email{vidybida@bitp.kiev.ua\\
http://www.bitp.kiev.ua/pers/vidybida}}

\date{Received: date / Revised: date}

\maketitle

\begin{abstract}
The binding neuron model \citep{Vid3} is inspired by numerical simulation of Hodg\-kin-Hux\-ley-type point neuron \citep{Vid}, as well as by the leaky in\-teg\-rate-and-fire model \citep{Seg}. In the binding neuron, the trace of an input is remembered for a fixed period of time after which it disappears completely. This is in the contrast with the above two models, where the postsynaptic potentials decay exponentially and can be forgotten only after triggering. The finiteness of memory in the binding neuron allows one to construct fast recurrent networks for computer modeling \citep{Vid1}. Recently, \citep{Vid4}, the finiteness is utilized for exact mathematical description of the output sto\-c\-h\-astic process if the binding neuron is driven with the Poisson input stream. In this paper, the simplest possible networking is considered for binding neuron. Namely, it is expected that every output spike of single neuron is immediately fed into its input. For this construction, externally fed with Poisson stream, the output stream is characterized in terms of interspike interval probability density distribution if the neuron has threshold 2. For higher thresholds, the distribution is calculated numerically. The distributions are compared with those found for binding neuron without feedback, and for leaky integrator. It is concluded that the feedback presence can radically alter spiking statistics.\\
\keywords{binding neuron \and feedback \and Poisson process \and interspike interval \and probability density \and information condensation}
\end{abstract}

\section{Introduction}

The main function of a neuron is to receive signals and to send them out. In real neurons, this function is realized through concrete biophysical mechanism, the main parts of which are ion channels in excitable membrane and variations of ionic concentrations inside and outside of nerve cell and its processes, see \citep{Schm} for details. The same function might be realized by means of any other mechanism able to support signals processing in the manner, which is characteristic of a real neuron. If so, then it would be interesting to develop a model, which realizes in an abstract form a concept of signal processing in real neurons, and is exempted from necessity to follow any biophysical mechanism supporting the processing. Such a model is necessary for quantitative mathematical formulation of what is going during signals/information processing in neural systems, see \citep{van} for discussion.  Attempts to develop such a model are mainly concentrated around concepts of coincidence detector and temporal integrator, see discussion in \citep{KonigTINS}. One more model, the binding neuron (BN), is proposed in \citep{Vid3}. This model is inspired by numerical simulation of Hodg\-kin-Hux\-ley-type  neuron stimulated from many synaptic inputs \citep{Vid}, as well as by the leaky integrate-and-fire model \citep{Seg}. It describes functioning of a neuron in terms of events, which are input and output spikes, and degree of temporal coherence between the input events, see \citep{Vid3,Vid4} for details.

It is observed, that during processing of sensory signals, the spiking statistics of individual neurons changes substantially when the signal travels from periphery to more central areas (see, e.g. \citep{Egger}). The changing of spiking statistics could underlie the information condensation, which happens during perception \citep{Konig}. This transformation of statistics may happen due to feedforward and feedback connections between neurons involved in the processing. Having in mind such possibilities, it would be interesting to check what happens with spike train statistical properties when it passes neuronal structures with feedback connections. 

\begin{figure}
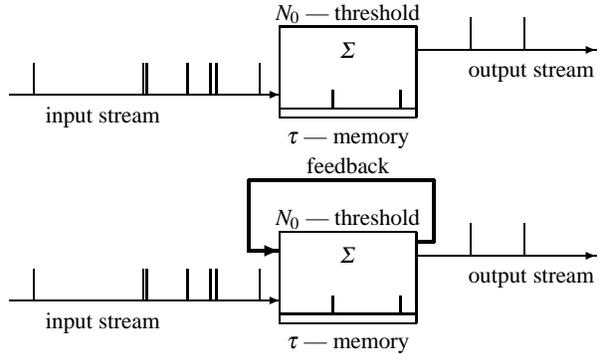

\unitlength=0.6mm
\begin{picture}(100,80)(-5,0)
\vbox{\hbox{\input{BN_en.pic}}\hbox{\input{BNB_en.pic}}}
\end{picture}
\caption{\label{BNB}
Schematic representation of binding neuron (top) and BN with feedback (bottom).}
\end{figure}

Usually, feedback/recurrent connections are considered between several neurons.
In this paper we consider the simplest possibility, namely, the single neuron with feedback. Such a configuration, which we regard as the simplest possible networking, can be found in real biological objects (see, e.g. \citep{Aron,Nicoll}).
As neuronal model we use binding neuron as it allows to obtain exact mathematical expressions suitable for further analysis.
 It is expected that input stream in any synapse of the neuron
 is Poisson one. In this case, from mathematical point of view, all inputs can be replaced with a single one with Poisson stream in it, having its intensity equal to the sum of all intensities in the synapses (Fig.\ref{BNB}, top). The binding neuron works as follows. Any input impulse is stored in the neuron during time $\tau$ and then it is forgotten. When the number of stored impulses, $\Sigma$, becomes equal to, or larger then the threshold one, $N_0$, the neuron sends an output spike, clears its internal memory and is ready to receive impulses from the input stream.
One obtains the binding neuron with feedback (BNF) by immediate feeding each output impulse to the neuron's input (Fig.\ref{BNB}, bottom). In this case, just after firing, the neuron has one impulse in its internal memory, and this impulse has time to live equal $\tau$. 

The specifics of mathematical analysis of BN-type systems is due to
presence in those systems both deterministic and stochastic dynamics.
Namely, the neuron obtains its input from a random stream (stochastic component) and every impulse is stored for the same fixed period of time (deterministic component).
This is in the contrast with the mass service theory \citep{Khi}, where the service time (counterpart of time to live, $\tau$) is random, Poisson-distributed. 
The simultaneous presence of deterministic and random dynamics in real neurons is due to the fact that in real neurons the impulse existence 
in a neuron (exposed as the excitatory postsynaptic potential) is supported by electrochemical transient \citep{H-H}, which is deterministic, whereas the input impulses come from other neurons and external media in irregular (random) manner\footnote{Compare with \cite{Goel}, \cite{brit}, where cases consistent with mass service theory are analyzed.}.

\section{Condensing of Information with Neurons}

It is widely accepted that during flow of sensory signals in a hierarchical manner from sensory periphery to central brain areas, the information, which is present in the signals, becomes less analogue and more discrete, eventually resulting in representing discrete symbols or entities (see e.g. \citep{Konig}). During this process, the amount of information within the flow must decrease in order to map various input spike trains from the sensory periphery into the same discrete entity. This process of consecutive reduction of information is known as condensation. We now put a question: What could be the primary element in which the condensation takes place? 
It seems that single neuron is a suitable candidate for such an element. In the case of binding neuron this can be explained as follows.

\begin{figure}
\begin{center}
\includegraphics[width=0.48\textwidth]{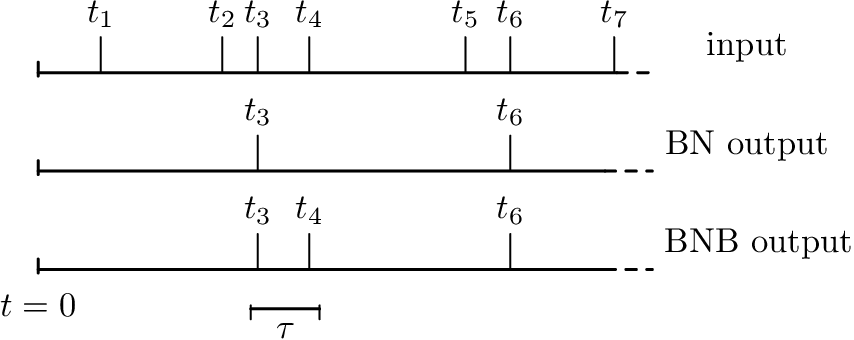}
\end{center}
\caption{\label{trains}
Example of input spike train and corresponding output for binding neuron (BN), and binding neuron with feedback (BNF). In both cases $N_0=2$.}
\end{figure}

Consider an input spike train like upper train in Fig.\ref{trains}. The train can be regarded as signal from a receptor neuron. After processing with the BN, the output spike train consists of two output impulses at moments $t_3$ and $t_6$. The BNF gives three spikes at moments $t_3$, $t_4$ and $t_6$. It is clear that the output trains contain less information then the input one. Nevertheless, having the output train for either BN, or BNF, one can make some condensed conclusions concerning the input train. 

In the case of BN, the output spike at moment $t_3$ tells us that there where input impulses at moments $t_3$ and $t^*$, where $t^*\in ]t_3-\tau;t_3[$, and the input impulse at moment $t^*$ did not trigger an output one at the moment of its arrival. For realization shown in Fig. \ref{trains}, $t^*=t_2$. Information delivered in the output spike $t_3$ is indeed condensed, because the presence and exact timing of this output spike does not change if position of $t^*$ deviates remaining within interval $]t_3-\tau;t_3[$. The same is about output spike at moment $t_6$.

In the case of BNF, the output spike at moment $t_3$ tells us that there where input impulses at moments $t_3$ and $t^*$, where $t^*\in ]t_3-\tau;t_3[$, and the time interval $]t^*-\tau;t^*[$ is free of input impulses. The same is about $t_6$. Two output spikes at moments $t_3$, $t_4$, where $t_4-t_3<\tau$, tell us additionally that in the input there were 3 consecutive impulses separated by intervals shorter then $\tau$.
Similar conclusions can be made for binding neurons with $N_0>2$. Moreover, condensed conclusion about input, which is based on corresponding output spike train, can be formulated for other neuronal models, like Hodgkin and Huxley, or leaky integrator. The difference is that for binding neuron the conclusion admits formulation in precise and clear mathematical manner, whereas for other models it does not. A less precise, fuzzy formulation, which is suitable for any model is that the output spike signals about presence in the input train temporally coherent (distributed over short time interval) sets of impulses (see \citep{Vid,Vid3} for discussion).

\section{Output Intensity of BNF with Threshold 2}

The intensity of output stochastic process can be defined by three different ways:
\begin{enumerate}
\item
It is possible to define the instantaneous intensity  (see \citep{Khi}), $\lambda_o(t)$, as the probability to obtain an output impulse at moment $t$ in infinitesimal interval $s$ divided by $s$:
\begin{equation}\label{laot}
\lambda_o(t)=\lim_{s\to0}\frac{w(s,t)}{s}\,,
\end{equation}
where $w(s,t)$ denotes the probability to obtain impulse in the interval
 $[t;t+s[$. 
We do not intend to analyze the $\lambda_o(t)$ time dependence here.
\item
 As limit, or mean output intensity one can choose the following
\begin{equation}\label{lao}
\lambda_o=\lim_{t\to\infty}\lambda_o(t)\,.
\end{equation}
\item
In the definition of $\lambda_o$ in (\ref{lao}), the role of time limit is to ensure that the initial state of the system is forgotten. In this case one can define/calculate $\lambda_o$ as the factor in the expression $\lambda_o\,dt$, which gives the probability to obtain an output impulse in the infinitesimal interval $dt$, if nothing is known about previous states of the neuron.
\end{enumerate}
It can be shown that both definitions 2 and 3 bring about the same value for $\lambda_o$. Therefore, we choose the third one here. Calculations based on the second definition can be fulfilled with the help of \cite[Part XI, \S8]{Feller}.

The probability to obtain an output impulse from the BNF with threshold 2 in the interval $dt$,  $\lambda_o\,dt$, can be calculated as product of probabilities of two independent events: (i) an input impulse is present in the  $dt$; (ii) the interval between that impulse and its predecessor is not longer than $\tau$. If the input stream is Poissonian, then the probability of event (i) is $\lambda\,dt$, and of event (ii) is $1-e^{-\lambda\,\tau}$, where $\lambda$ is the intensity of input stream. Thus,
\begin{equation}\label{laou}
\lambda_o=(1-e^{-\lambda\,\tau})\lambda\,.
\end{equation}

\section[Distribution of Output Intervals for BNF with $N_0=2$]{Distribution of Output Intervals for BNF with $N_0=2$}

Let us consider a BNF with threshold $N_0=2$ and internal memory $\tau$, which obtains its input from Poisson stream with intensity $\lambda$.
Thus, the neuron fires every time when input impulse comes $\tau$, or less units of time after its predecessor.

The output statistics can be described in terms of the probability density distribution to obtain an output interspike interval $t$ with precision $dt$. For this purpose it is enough to calculate the probability, $P_b(t,\tau)dt$, of the following event: the next firing happens $t$ units of time later than the previous one. Let the input impulses, which come after the previous firing, are numbered with numbers 1, 2, \dots .

The above-mentioned event can be decomposed into several alternatives, which are numbered with the number $k$ of input impulse, which triggers the next firing. Notice, that for $t<\tau$ only one alternative is possible. It happens if the first input impulse comes not later then $\tau$ units of time after the previous firing. In this case, the neuron still keeps impulse from the previous firing, and the input causes the threshold 
achieving and firing. There is no other ways to get output interval $t<\tau$. Thus, for $t\in[0;\tau[$, the probability density distribution is as follows
\begin{equation}\label{Pb0}
P_b(t,\tau)\,dt=e^{-\lambda\,t}\lambda\,dt\,.
\end{equation}

It is impossible to obtain output interval $t>\tau$ with a single input impulse.\footnote{The value of $P_b(\tau,\tau)$ can be chosen arbitrary.} Thus, for $t>\tau$, possible alternatives are numbered with numbers 2, 3, \dots, $k_{max}$, where
$
k_{max}=\left[{t/\tau}\right]+1\,,
$
and $[x]$ denotes the integral part of $x$.  

Assume, the $k$-th alternative is realized by input arrival times $t_1, t_2, \dots, t_{k-1}$, $t_k\equiv t$. Not all arrival times are admitted (see (\ref{umovy2}) and further, below). In accordance with the definition of Poisson process, the probability of such realization is given by the following expression:
$$
e^{-\lambda t_1}\lambda dt_1
e^{-\lambda(t_2-t_1)}\lambda dt_2
\cdots
e^{-\lambda(t-t_{k-1})}\lambda dt\,.
$$
The probability $P_{bk}(t,\tau)dt$ that the $k$-th alternative is realized with any admissible values of $t_1,$ $t_2,$ $\dots$, $t_{k-1}$ can be calculated by integrating of the above expression over the region of $(k-1)$-dimensional space with coordinates
$t_1$, $t_2, \dots,$ $t_{k-1}$, defined by the following conditions:
\begin{equation}\label{umovy2}
t_1\ge\tau,~~t_1+\tau<t_2,~~ \dots,~~
t_{k-2}+\tau<t_{k-1}<t\,,
\end{equation}
and $t-t_{k-1}<\tau$.
The required integral over the region defined by (\ref{umovy2}) can be calculated exactly:
\begin{multline}\label{Pk3}
e^{-\lambda t}\lambda^{k-1}
\int\limits_{\tau}^{t-(k-2)\tau} dt_1
\int\limits_{t_1+\tau}^{t-(k-3)\tau} dt_2
\cdots
\int\limits_{t_{k-2}+\tau}^{t} dt_{k-1}
\lambda dt=\\
=e^{-\lambda t}\lambda^{k-1}{(t-(k-1)\tau)^{k-1}\over(k-1)!}
\lambda dt\,.
\end{multline}
If $k=k_{max}$, then (\ref{umovy2}) ensures: $(k-1)$-th impulse
is in the interval $]t-\tau;t[$, and $k$-th impulse at moment $t$ will
cause firing. Thus, in this case
$$
P_{b\,k}(t,\tau)\,dt=
e^{-\lambda t}\lambda^{k-1}{(t-(k-1)\tau)^{k-1}\over(k-1)!}
\lambda dt\,,\quad
k=k_{max}\,.
$$
If $k<k_{max}$, then integral (\ref{Pk3}) includes also configurations for which $t_{k-1}<t-\tau$. For these configurations $k$-th input impulse at moment $t$ will not cause firing. The contribution of these configurations into the integral (\ref{Pk3}) is given by the following expression
\begin{multline*}
e^{-\lambda t}\lambda^{k-1}
\int\limits_{\tau}^{t-(k-1)\tau} dt_1
\int\limits_{t_1+\tau}^{t-(k-2)\tau} dt_2
\cdots
\int\limits_{t_{k-2}+\tau}^{t-\tau} dt_{k-1}
\lambda dt=
\\
=e^{-\lambda t}\lambda^{k-1}{(t-k\,\tau)^{k-1}\over(k-1)!}
\lambda dt\,,
\end{multline*}
which should be subtracted from (\ref{Pk3}). Thus, for $2\le k<k_{max}$:
\begin{multline*}
P_{b\,k}(t,\tau)\,dt =
\\
=e^{-\lambda t}{\lambda^{k-1}\over(k-1)!}
\left((t-(k-1)\tau)^{k-1} - (t-k\,\tau)^{k-1}\right)
\lambda dt\,.
\end{multline*}
\begin{figure}
\begin{center}
\includegraphics[width=0.16\textwidth,angle=-90]{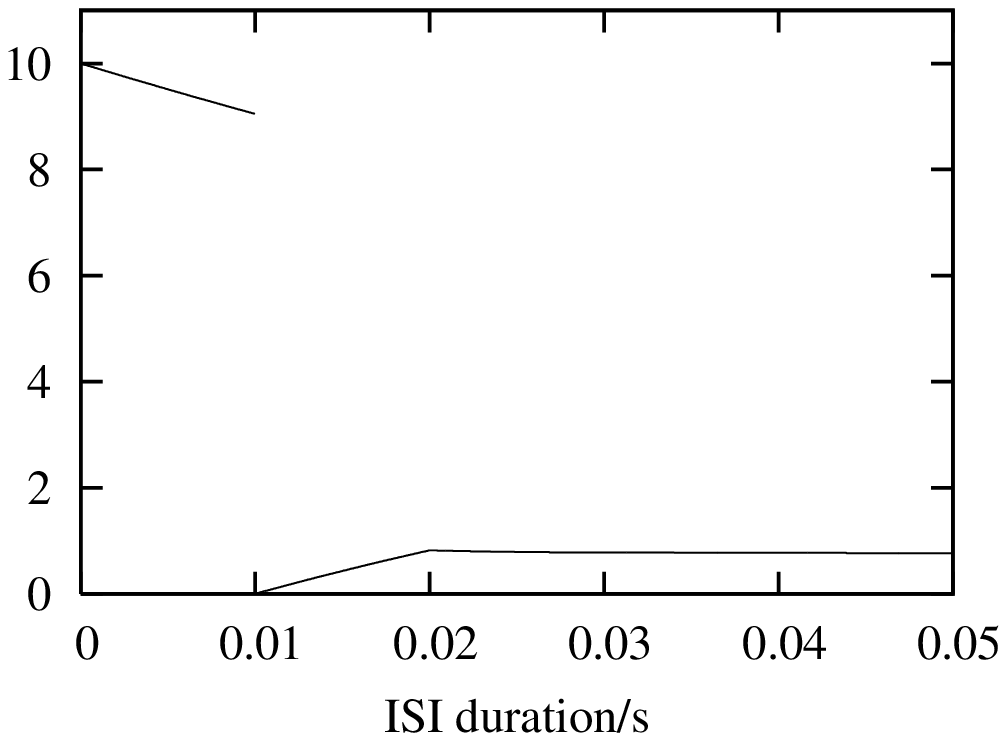}
\includegraphics[width=0.16\textwidth,angle=-90]{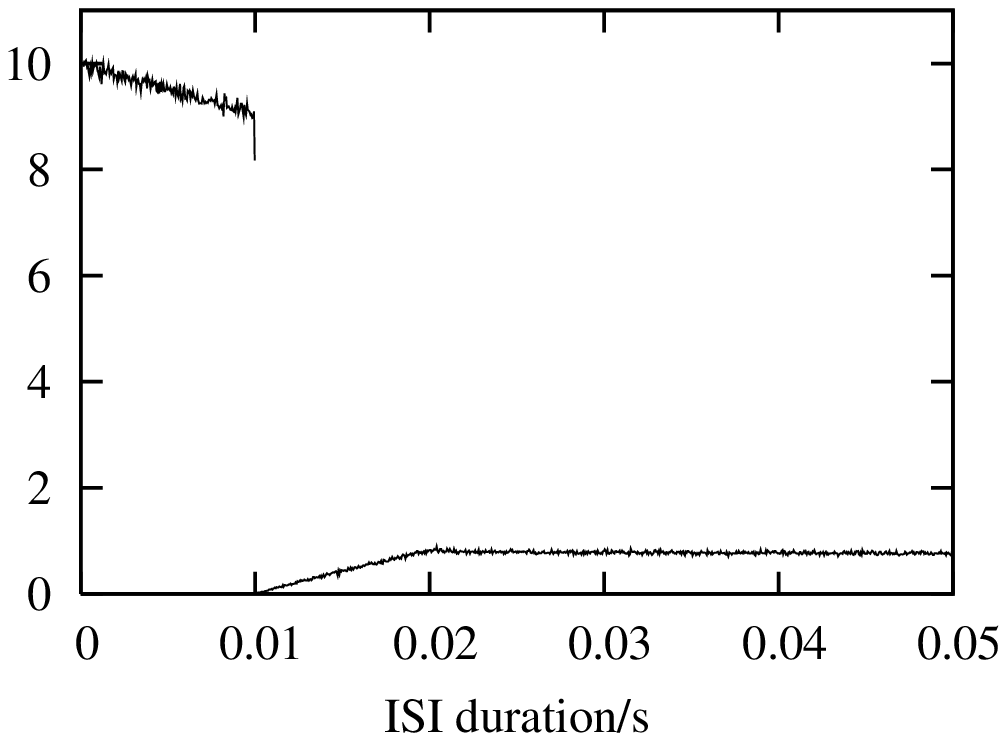}
\end{center}
\caption{\label{la_b}
Interspike intervals (ISI) distribution $P_b(t,\tau)$ for $\tau$ = 10 ms, $\lambda$ = 0.01 ms$^{-1}$, $N_0=2$. Left --- calculated in accordance with (\ref{Pb0}), (\ref{mtaub}), right --- calculated numerically.}
\end{figure}
The total probability is calculated by summation over all alternatives.
Notice, that $k_{max}$ changes by 1 when $t$ passes through integer multiple of $\tau$. Thus, for $m=1,2,\dots$ the following statement is valid: if $m\tau<t<(m+1)\tau$, then
\begin{equation}\label{mtaub}
P_b(t,\tau)dt=
e^{-\lambda t}{\lambda^m\over m!}(t-m\tau)^m
\lambda dt+
\end{equation}
$$+
\sum\limits_{2\le k\le m}
e^{-\lambda t}{\lambda^{k-1}\over(k-1)!}
\left((t-(k-1)\tau)^{k-1} - (t-k\,\tau)^{k-1}\right)
\lambda dt\,.
$$
For $t\in[0;\tau[$ the function $P_b(t,\tau)\,dt$ is given by (\ref{Pb0}).
The distribution $P_b(t,\tau)\,dt$ is analogous to distribution $P(t)\,dt=e^{-\lambda t}\lambda\,dt$ known for Poisson process. The graph of $P_b(t,\tau)$ is shown in Fig.\ref{la_b}.

\section{Properties of the distribution}

Notice that after firing, the neuron starts from standard state: it keeps a single impulse with time to live equal $\tau$. Therefore, there is no correlation between consecutive interspike intervals.

\subsection{Connection with BN distribution}
It is interesting that function $P(t,\tau)$, which gives the probability density distribution for binding neuron without feedback (see \cite[Eq. (6)]{Vid4}) has simple interconnection with $P_b(t,\tau)$. In order to find this interconnection, denote restriction of $P_b(t,\tau)$ onto interval $[m\tau;(m+1)\tau[$ as $P_{bm}(t,\tau)$.
(\ref{mtaub}) then means:
$$m\tau\le t<(m+1)\tau ~\Rightarrow~ P_b(t,\tau)=P_{bm}(t,\tau),~
m=1,2,\dots\,.$$
Substitute here $t+\tau$ instead of $t$:
$$m\tau\le t+\tau<(m+1)\tau ~\Rightarrow~ P_b(t+\tau,\tau)=P_{bm}(t+\tau,\tau),$$
where $m=1,2,\dots\,,$ or
$$(m-1)\tau\le t<m\tau ~\Rightarrow~ P_b(t+\tau,\tau)=P_{bm}(t+\tau,\tau),$$
where  $m=1,2,\dots\,.$
Substitute here $m$ instead of $(m-1)$:
\begin{multline*}
m\tau\le t<(m+1)\tau ~\Rightarrow~ P_b(t+\tau,\tau)=P_{b,m+1}(t+\tau,\tau),\\ m=0,1,2,\dots\,.
\end{multline*}
The explicit expression for $P_{b,m+1}(t+\tau,\tau)$ can be obtained from (\ref{mtaub}):
\begin{multline*}
P_{b,m+1}(t+\tau,\tau)=
e^{-\lambda (t+\tau)}{\lambda^{m+1}\over (m+1)!}(t-m\tau)^{m+1}
\lambda +
\\+
\sum\limits_{2\le k\le m+1}
e^{-\lambda (t+\tau)}{\lambda^{k-1}\over(k-1)!}
\big((t-(k-2)\tau)^{k-1} -
\\-
(t-(k-1)\tau)^{k-1}\big)
\lambda\,.
\end{multline*}
The last expression coincides with the corresponding term in the \cite[Eq. (6)]{Vid4} multiplied by $e^{-\lambda\tau}$. 
Thus, the following representation takes place:
\begin{equation}\label{intercon}
\begin{cases}
0\le t<\tau ~\Rightarrow~ P_b(t,\tau)=e^{-\lambda t}\lambda,\\
\tau\le t ~\Rightarrow~ P_b(t,\tau)=e^{-\lambda\tau}P(t-\tau,\tau)\,.
\end{cases}
\end{equation}
The last expression together with the fact that $P(t,\tau)$ from \cite[Eq (6)]{Vid4} is normalized, allows one to check easily that $P_b(t,\tau)$ is normalized as well:
$$
\int\limits_0^\infty\,P_b(t,\tau)\,dt=1.
$$

\subsection{Mean interspike interval}

Having for $P_b(t,\tau)$ representation (\ref{intercon}), one can easily calculate mean interspike interval, $W$, which is defined as
$$
W=\int\limits_0^\infty\,t\,P_b(t)\,dt.
$$
Substitute here representation (\ref{intercon}):
\begin{multline*}
W=\int\limits_0^\tau\,t\,e^{-\lambda t}\lambda\,dt+
\int\limits_\tau^\infty\,t\,e^{-\lambda\tau}P(t-\tau,\tau)\,dt=
\\
=\frac{1-e^{-\lambda\tau}}{\lambda} - \tau\,e^{-\lambda\tau}+
e^{-\lambda\tau}\,\int\limits_0^\infty\,(t+\tau)\,P(t,\tau)\,dt=
\\
=\frac{1-e^{-\lambda\tau}}{\lambda} +
e^{-\lambda\tau}\,\int\limits_0^\infty\,t\,P(t,\tau)\,dt.
\end{multline*}
The last integral is calculated in \cite[Sec. 3.2]{Vid4}. Use found there expression:
\begin{multline}
W=\frac{1-e^{-\lambda\tau}}{\lambda} +
e^{-\lambda\tau}\,\frac{1}{\lambda}\left(2+\frac{1}{e^{\lambda\tau}-1}\right)=
\\\label{W}
=\frac{1}{\lambda\left(1-e^{-\lambda\tau}\right)}\,.
\end{multline}

\subsection{Coefficient of variation}\label{CVS}

Coefficient of variation $c_{bv}$ for obtained distribution (\ref{mtaub}) can be calculated as follows
\begin{equation}\label{CVdef}
c_{bv}=\sqrt{\frac{W_2}{W^2}-1}\,,
\end{equation}
where $W$ is given in (\ref{W}), and $W_2$ is the second moment of distribution  (\ref{mtaub}):
$$
W_2=\int\limits_0^\infty\,t^2\,P_b(t,\tau)\;dt.
$$
Here use representation (\ref{intercon}):
\begin{equation}\label{W_2}
W_2=\lambda\,\int\limits_0^\tau\,t^2\,e^{-\lambda t}\;dt+
e^{-\lambda\tau}\,\int\limits_\tau^\infty\,t^2 P(t-\tau,\tau)\;dt=
\end{equation}
$$
=\frac{2-((\lambda\tau)^2+2\lambda\tau+2)e^{-\lambda\tau}}{\lambda^2}+
e^{-\lambda\tau}\,\int\limits_0^\infty\,(t+\tau)^2 P(t,\tau)\;dt.
$$
The second term here can be split into three:
$$
e^{-\lambda\tau}\,\int\limits_0^\infty\,\tau^2 P(t,\tau)\;dt=e^{-\lambda\tau}\,\tau^2,
$$
$$
e^{-\lambda\tau}\,2\tau\,\int\limits_0^\infty\,t P(t,\tau)\;dt=
e^{-\lambda\tau}\,2\tau\,\frac{1}{\lambda}\left(2+\frac{1}{e^{\lambda\tau}-1}\right),
$$
(used same expression as for calculating (\ref{W})), and
$$
e^{-\lambda\tau}\,\int\limits_0^\infty\,t^2 P(t,\tau)\;dt.
$$
The ISI distribution's second moment for BN without feedback can be calculated similarly as it is done for its first moment. This gives
\begin{equation}\label{BNW_2}
\int\limits_0^\infty\,t^2 P(t,\tau)\;dt=
\frac{2}{\lambda^2}\,\frac{3\,e^{2\,\lambda\,\tau}+(\lambda\,\tau-3)\,e^{\lambda\,\tau}+1}{(e^{\lambda\,\tau}-1)^2}.
\end{equation}
Substitute this into (\ref{W_2}). This gives 
\begin{equation}\label{W_2fin}
W_2=\frac{2\,e^{\lambda\,\tau}}{\lambda^2}\,\frac{e^{\lambda\,\tau}+\lambda\,\tau}{\left(e^{\lambda\,\tau}-1\right)^2}\,.
\end{equation}
Substitute this and (\ref{W}) into (\ref{CVdef}), this gives
$$
c_{bv}=\sqrt{2\,\lambda\tau\,e^{-\lambda\tau}+1}\,.
$$
Coefficient of variation gets its maximum value, $c_{bvm}$,
$$c_{bvm}= \sqrt{2\,e^{-1}+1}\approx1.32$$ at $\lambda\tau=1$ (Fig. \ref{Th2BN}).

It is also possible, by using Eq. (\ref{BNW_2}), to calculate coefficient of variation, $c_v$, for BN without feedback:
$$
c_{v}=\sqrt{\frac{2\,\lambda\tau\,e^{\lambda\tau}+0.5}{4\,e^{2\lambda\tau}- 4\,e^{\lambda\tau}+1}+\frac{1}{2}}\,.
$$

The $c_v$ gets its maximum value equal to 1 at $\lambda\tau=0$, and decreases monotonically when $\lambda\tau$ increases (Fig. \ref{Th2BN}). 

\section{Numerical Simulations}\label{num}

\begin{figure}
\begin{center}
\includegraphics[width=0.16\textwidth,angle=-90]{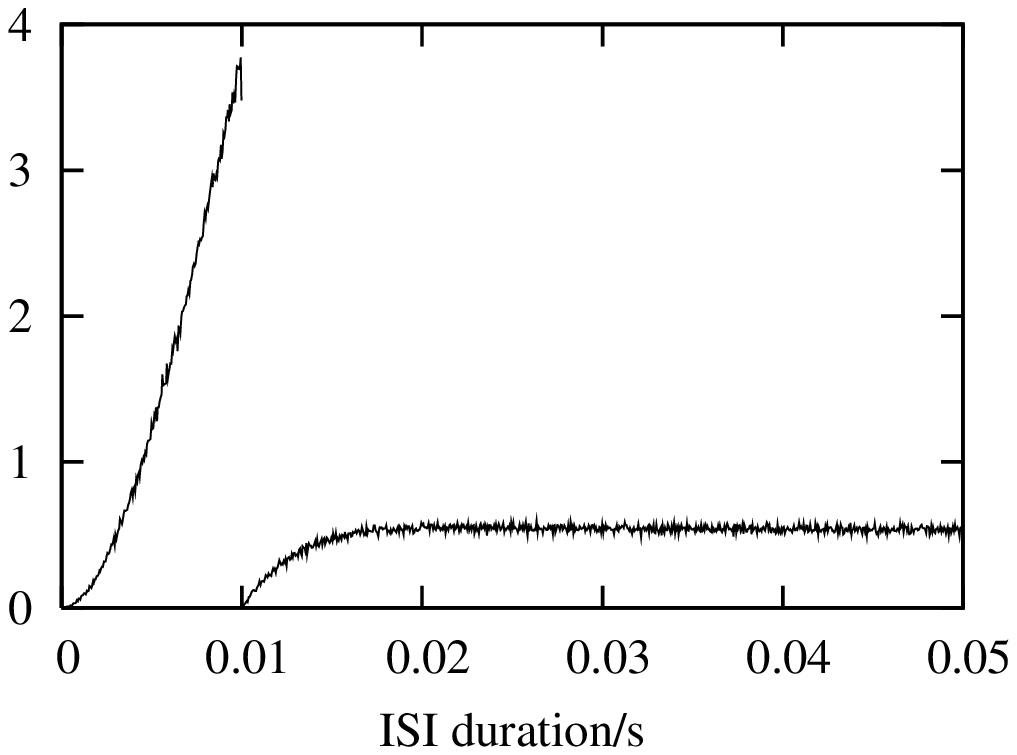}
\includegraphics[width=0.16\textwidth,angle=-90]{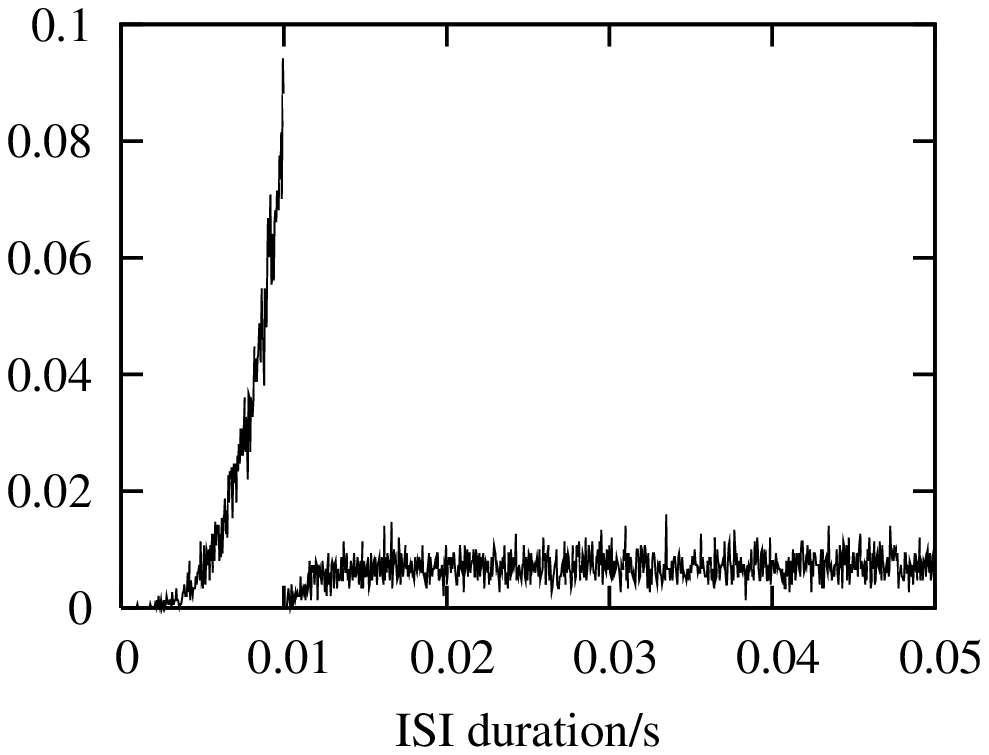}
\end{center}
\caption{\label{Th46}ISI distribution $P_b(t,\tau)$ found numerically for $\tau$ = 10 ms, $\lambda$ = 0.05 ms$^{-1}$. Left --- $N_0=4$, right --- $N_0=6$. Notice the discontinuity at $t=\tau$. $30\,000\,000$ triggerings were taken in both cases.}
\end{figure}

Numerical simulations were executed here for several purposes. The first purpose was to check numerically correctness of the expressions found analytically in previous sections. A C$^{++}$ program was developed, which allows to calculate the $P_b(t,\tau)$. The Poisson streams were generated by transformation of uniformly distributed sequences of random numbers (see, e.g. Eq. (12.14) in  \citep{Cell}). Those sequences were produced with the system pseudorandom number generator from \verb+libc+ library in the Linux operating system, as well as with the Mersenne Twister, \verb+mt19937+ \citep{Mat}. The two methods give indistinguishable results.  The program includes the BNF class, which analyzes the input stream and fires in accordance with the rules, described above. With the help of that class, output stream samples were produced by calculating $N=30\,000\,000$ output spikes. The samples are scanned for interspike intervals of various duration, and the probability density distribution is then calculated by normalization. The numerically obtained ISI distributions are in good agreement with analytical expression, as it can be seen in Fig. \ref{la_b}. Also, the second moment of $P_b(t,\tau)$ was calculated numerically for several values of $\lambda$, $\tau$. Deviation of numerically found values from analytical expression (\ref{W_2fin}) is within 0.01\% $\sim$ 0.1\% range.

The second purpose of numerical calculations is to obtain ISI distributions for higher thresholds.
The above mathematical reasoning for BNF with threshold $N_0=2$ becomes extremely cumbersome for thresholds $N_0=3$, or higher. It seems that there should be a more effective mathematical approach for higher thresholds. Meanwhile, it is possible to calculate numerically the probability density distribution for any threshold value. Examples of calculated densities are shown in Fig. \ref{Th46}. These densities are in qualitative agreement with what is found analytically for $N_0=2$, except of the fact that the initial part of ISI distribution is increasing for $N_0>2$, whereas for $N_0=2$ it is decreasing. 
The initial (for $t<\tau$) part of the probability density distribution $P_b(t,\tau)dt$ can be easily found analytically for any threshold $N_0$. Indeed, denote the moment of the previous firing as 0. At this moment BNF stores one impulse with time to live $\tau$. The next firing happens at moment $t<\tau$ iff $N_0-2$ input impulses come within the interval $]0;t[$, and one more impulse within $[t;t+dt[$. The probability of such event for Poisson process is known, which gives for any $N_0\ge2$
$$
P_b(t,\tau)\,dt=e^{-\lambda t}\frac{(\lambda t)^{N_0-2}}{(N_0-2)!}\lambda dt,\quad t<\tau\,.
$$
This function is decreasing for $N_0=2$ and increasing for higher $N_0$, which
explains seeming qualitative disagreement between $N_0=2$ and $N_0>2$ cases.

The third purpose was to compare the ISI distributions found here for the binding neuron model with those for leaky integrate and fire (LIF) model. In the program developed, the BNF class was replaced with LIF class, which reproduces the simplest version of the LIF model. Namely, the LIF neuron is characterized by a threshold, $C$, and every input impulse advances by $y_0$ the LIF membrane voltage, $V$. Between input impulses, $V$ decays exponentially with time constant $\tau_M$. The LIF neuron fires when $V$ becomes greater or equal $C$, and $V=0$ just after firing. Examples of the ISI distribution obtained for various parameter values are shown in Fig. \ref{LIF}.

\section{Conclusions}

\begin{figure}[h]
\begin{center}
\includegraphics[width=0.16\textwidth,angle=-90]{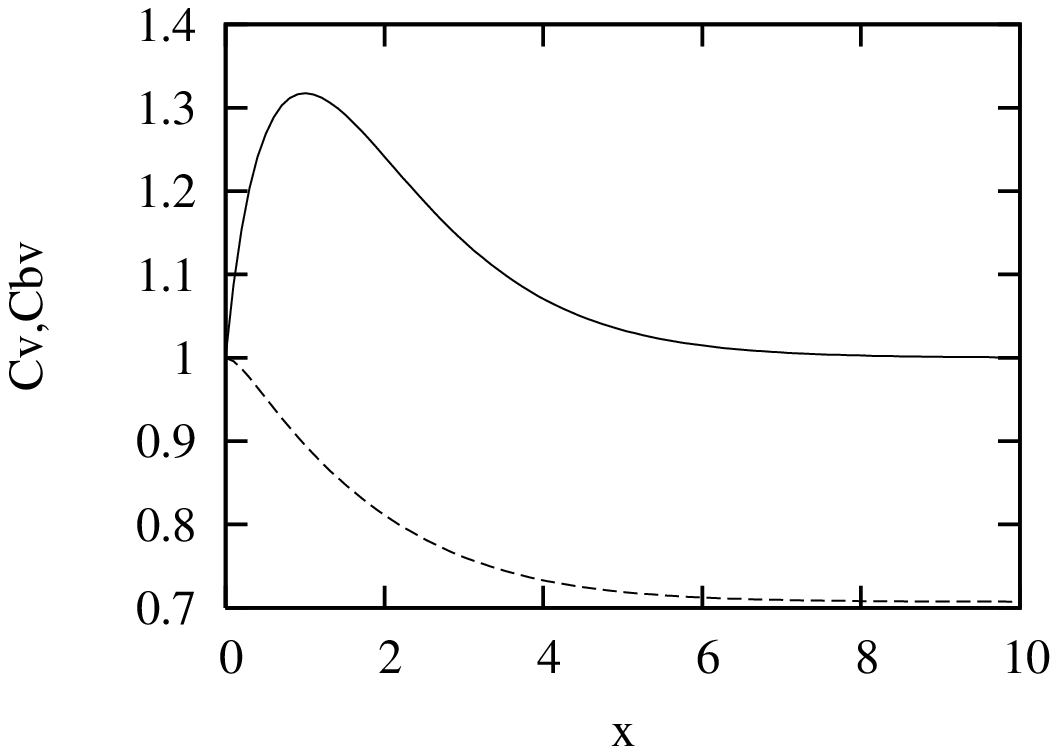}
\includegraphics[width=0.16\textwidth,angle=-90]{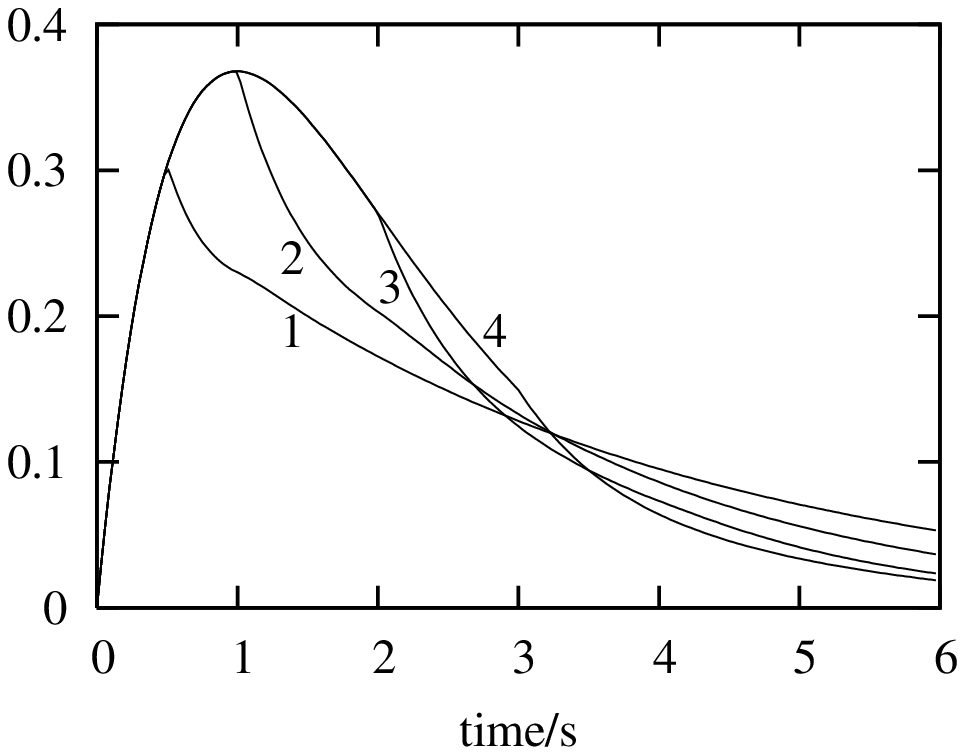}
\end{center}
\caption{\label{Th2BN}Left: Coefficient of variation as function of $x=\lambda\tau$ for BN (lower curve) and BNF (upper curve). Right: Interspike intervals distribution $P(t,\tau)$ for BN with $N_0=2$ and without feedback (from \citep{Vid4}). Here $\lambda=1$ s$^{-1}$; curves number 1,2,3,4 correspond to $\tau$ = 0.5 s, 1 s, 2 s, 3 s.}
\end{figure}

We calculated here the intensity and output interspike intervals distribution for binding neuron with feedback, which is stimulated with Poisson stream. For BNF with threshold $N_0=2$ this is done analytically, for higher thresholds --- numerically. It is interesting to compare the obtained distributions with those known for other models. 
In Fig. \ref{Th2BN}, the distribution is shown for binding neuron without feedback. Curve 4 in Fig. \ref{Th2BN} is qualitatively similar to distribution obtained numerically in \citep{Seg} for leaky integrator model in a slightly different stimulation paradigm.
By comparing these distributions with those found here for BNF (see also comparison of coefficients of variation in sec. \ref{CVS}) , one could conclude that even the simplest possible networking is able to change radically statistical properties of spiking process. This gives a hint about what could take place with spiking statistics of individual neurons in a network.

Numerical calculations made for the LIF model (Fig. \ref{LIF}) suggest, that
 introducing feedback might result in qualitative changing of spiking statistics  for other neuronal models as well.

\section{Discussion}

The model of binding neuron used here is simplified in a sense, that it does not follow time course of ionic currents, or transmembrane voltage. The purpose of this model (see \citep{Vid3}) is to formulate in abstract form the answer to the question: What does neuron do with signals it receives? The question well can be answered in the framework of more detailed models, like \citep{H-H}. But usage of more detailed models for description of less basic functions, like neural coding, or information processing, would be the same as to describe computer functioning in terms of Kirchhoff's laws: it is correct, but not productive.

\begin{figure}
\begin{center}
\includegraphics[width=0.16\textwidth,angle=-90]{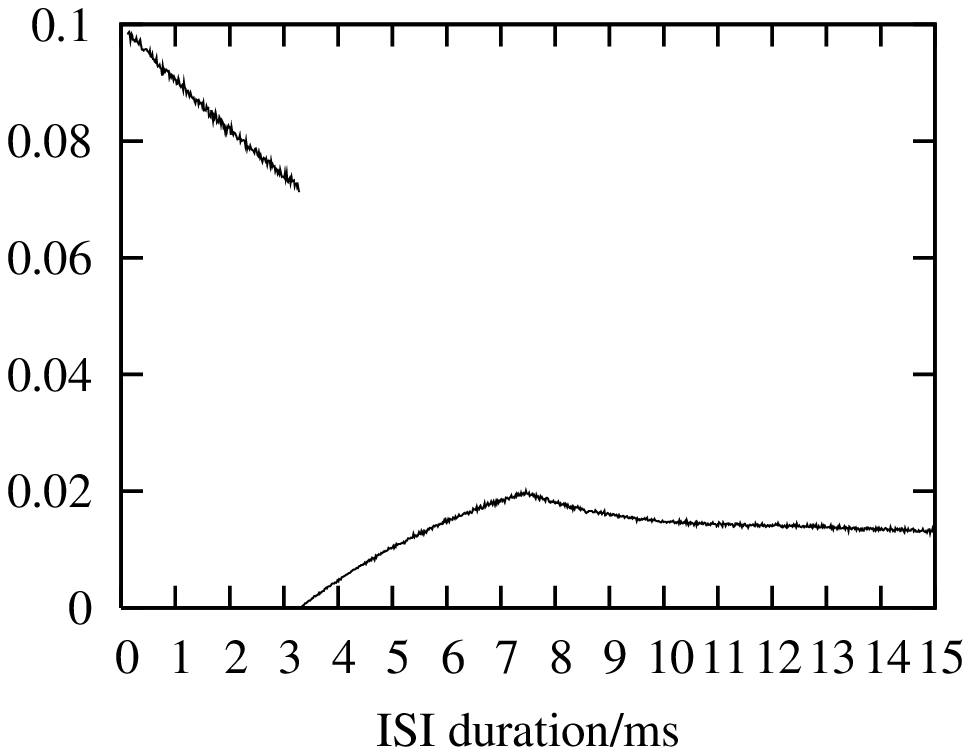}
\includegraphics[width=0.16\textwidth,angle=-90]{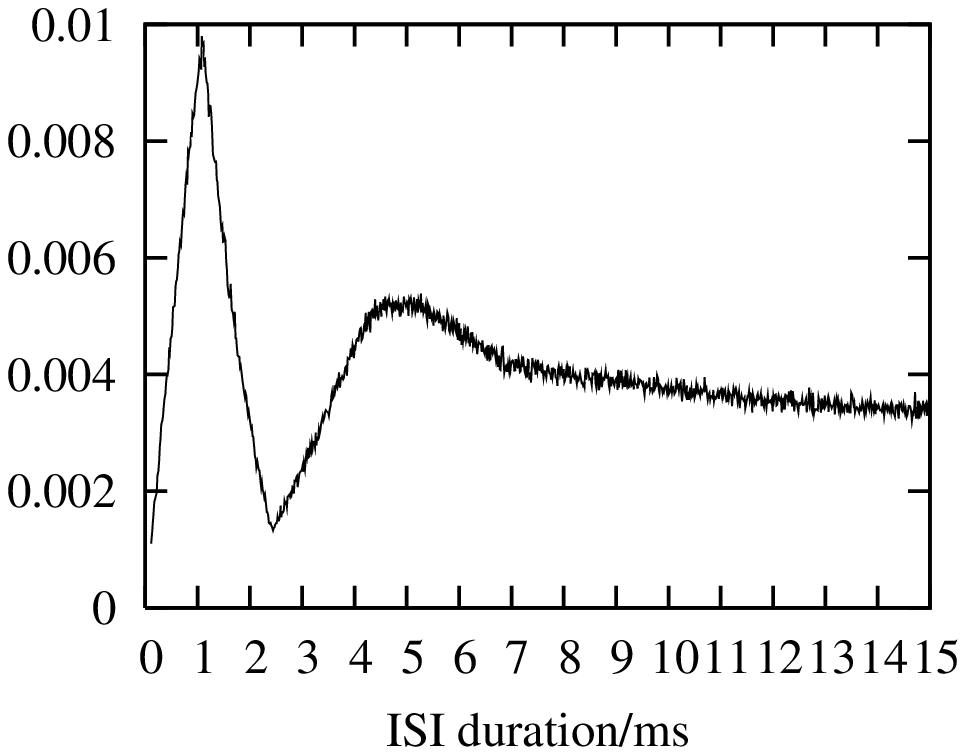}
\end{center}
\caption{\label{LIF}ISI distribution $P_b(t)$ found numerically for leaky integrate and fire model with feedback. Used  $30\,000\,000$ output spikes. Firing threshold, $C=20$ mV, input intensity, $\lambda=0.1$ ms$^{-1}$. Left: membrane time constant, $\tau_M = 3$ ms, input impulse amplitude, $y_0= 15$ mV. Right: $\tau_M=6$ ms , $y_0= 7.5$ mV.}
\end{figure}

The exact discontinuities in the output ISI distributions, which can be seen in Figs. \ref{la_b},\ref{Th46}, are due to abrupt loss of feedback input influence $\tau$ units of time after triggering. Output ISI, which is shorter then $\tau$, is created with the feedback spike involved. The longer ISIs are created without feedbacked spike involvement. Therefore, the jump is in the direction of smaller probabilities. In the models, in which the influence of input spike diminishes gradually, one could expect the decreasing region of probability density function course in the range, where role of feedback inputs becomes small. This could cause a bimodal distribution of output ISIs, like shown in Fig. \ref{LIF}, right. Nevertheless, for special parameter values, the genuine discontinuity can be as well observed for the LIF model, like in Fig. \ref{LIF}, left\footnote{For the LIF model, presence of discontinuity in ISI distribution in Fig. \ref{LIF}, left, can be proved mathematically rigorously.}.

The coefficients of variation dependence on $x=\lambda\tau$ can be explained as follows. For $x\to0$ both BN, and BNF output streams become Poissonian. Consider the BN case. The BN will generate an output spike in interval $[t;t+dt[$ if three conditions are satisfied: (i) there is input spike in $[t;t+dt[$, (ii) the previous input was received at $t-\tau$, or later, (iii) the previous input did not triggered BN. Violation of cond. (iii) with (i), (ii) satisfied is improbable when $\lambda\tau\to0$, because this means appearance of two consecutive input ISIs, both shorter then $\tau$. For Poisson input this may happen with probability $\left(1-e^{-\lambda\tau}\right)^2$, and for small $x$ may be neglected. In this case the desired probability of output is $\left(1-e^{-\lambda\tau}\right)\lambda\,dt$, which describes Poisson stream with intensity $\lambda'=\left(1-e^{-\lambda\tau}\right)\lambda$. For this stream, coefficient of variation will be 1. Similar reasoning are valid for BNF. 
In the opposite case, when $\lambda\tau\to\infty$, violation of condition (iii) for BN cannot be ignored. Actually, for high stimulation rates, the BN will act as perfect integrator. The output stream of perfect integrator is $\gamma$-distributed, with $c_v<1$. For BNF at high stimulation rates, every feedbacked spike will combine with next input one, and trigger next output spike\footnote{$N_0=2$ is expected.}. This possibility was mentioned as ``dancing in step'' in \citep[p.43]{MacKay}. In such a regime, output stream exactly reproduces the input one, hence, is Poisson stream with $c_{bv}=1$. For intermediate values of $\lambda\tau$ the ``dancing in step'' will be interrupted from time to time by waiting longer then $\tau$ for the next input spike. The triggering, which is next to this event, must happen without feedback involvement. Combination of this two possibilities gives maximum variability of output stream at $\lambda\tau=1$.

Finally, it would be interesting to compare ISI distributions found here with those observed experimentally. The configurations with feedback are known for real biological objects, \citep{Aron,Nicoll}. The self-excitating neurons described in the cited papers are incorporated in a complicated network. Thus, their spiking statistics is influenced by other neurons. Therefore, a more developed network model is needed in order to compare with experimental data. Such a model will be studied in future.

\vspace*{\baselineskip}\noindent
{\small{\bf Acknowledgments.} The author thanks to referee for stimulating suggestions. Also, I thank to A.~Andrew for sending me the D.~MacKay's paper. 
During preparation of this paper the following free software were used: (i) Linux operating system with accompanying staff, like \verb+gcc+, \verb+libc+, \verb+gnuplot+, \LaTeX; (ii) computer algebra system ``Maxima'' (http://maxima.sourceforge.net).}

\end{document}